\documentclass[letterpaper, 10 pt, conference]{ieeeconf}  

\IEEEoverridecommandlockouts

\usepackage{graphicx}
\usepackage{amsfonts}
\usepackage{amsmath}
\usepackage{amssymb}
\usepackage{color}

\usepackage{graphicx,color,overpic}
\usepackage{psfrag}
\usepackage{amssymb}
\usepackage{times}
\usepackage{latexsym}
\usepackage{bm}
\usepackage{cases}
\usepackage{array}
\usepackage{fancyhdr}
\usepackage{setspace}
\usepackage{subfigure}
\usepackage{url}
\usepackage{multirow}
\usepackage{epstopdf}

\usepackage{epsfig}
\usepackage{fancybox}
\usepackage{textcomp}

\newcommand{\Z}[0]{\mathcal{Z}}
\newcommand{\M}[0]{\mathcal{M}}
\newcommand{\x}[0]{\mathbf{x}}
\newcommand{\N}[0]{\mathcal{N}}
\newcommand{\z}[0]{\mathbf{z}}

\newcommand{\m}[0]{\mathbf{m}}

\IEEEoverridecommandlockouts                              

\title{\LARGE \bf
Cooperative Relative Positioning of Mobile Users by Fusing IMU Inertial and UWB Ranging Information
}

\author{Ran Liu, Chau Yuen, Tri-Nhut Do, Dewei Jiao, Xiang Liu, and U-Xuan Tan
\thanks{This work is supported by Temasek Lab under Indoor Relative Positioning System project (No. IGDST1302024), National Science Foundation of China (No. 61550110244, 61601381, and 61471306), 
and National Defense Scientific Research of China (No. B3120133002).} 
\thanks{R. Liu, C. Yuen, T. N. Do, and U-X. Tan are with the Engineering Product Development Pillar, Singapore University of Technology and Design, 8 Somapah Rd, Singapore, 487372 
{\{\tt\small ran\_liu, yuenchau, trinhut\_do, uxuan\_tan\}@sutd.edu.sg}.}
\thanks{D. Jiao and X. Liu are with the School of Software and Microelectronics, Peking University, Beijing, China, 102600 {\tt\small xliu@ss.pku.edu.cn}.
}}

\begin{document}

\maketitle
\thispagestyle{empty}
\pagestyle{empty}

\begin{abstract}
Relative positioning between multiple mobile users is essential for many applications, such as search and rescue in disaster areas or human social interaction.
Inertial-measurement unit (IMU) is promising to determine the change of position over short periods of time, but it is very sensitive to error accumulation over long term run.
By equipping the mobile users with ranging unit, e.g. ultra-wideband (UWB), it is possible to achieve accurate relative positioning by trilateration-based approaches.
As compared to vision or laser-based sensors, 
the UWB does not need to be with in line-of-sight and provides accurate distance estimation.
However, UWB does not provide any bearing information and the communication range is limited, 
thus UWB alone cannot determine the user location without any ambiguity.
In this paper, 
we propose an approach to combine IMU inertial and UWB ranging measurement for relative positioning between multiple mobile users without the knowledge of the infrastructure. 
We incorporate the UWB and the IMU measurement into a probabilistic-based framework, 
which allows to cooperatively position a group of mobile users and recover from positioning failures. 
We have conducted extensive experiments to demonstrate the benefits of incorporating IMU inertial and UWB ranging measurements.


\end{abstract}

\section{Introduction}
\label{Introduction}
Indoor positioning systems are essential to provide many public, commercial, and military services.
Many researchers concentrate on the absolute positioning in a global coordinate system with respect to a specific infrastructure, 
where many reference anchors with known positions are deployed. 
The user measures the received signal strength (RSS)\,\cite{Yassin_ieee_tutorials_2016}, time of arrival (ToA)\,\cite{time_of_arrival_2011}, or angle of arrival (AoA)\,\cite{angle_of_arrival_2006} 
to anchors and infers its position in the environment. 
A typical system is the global positioning system (GPS), 
which utilizes the satellites as anchors to provide position information in outdoor environments with an accuracy of several meters. 

\begin{figure}
\centering
\includegraphics[width=0.45\textwidth]{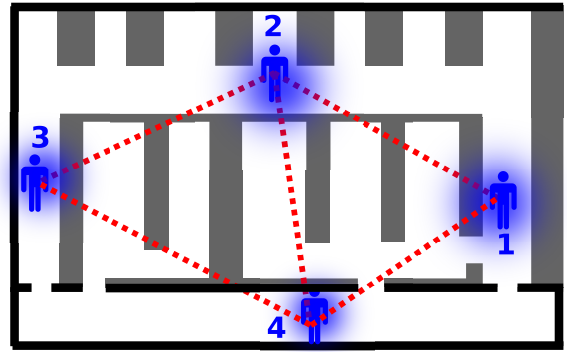}
\caption{
Illustration of the relative positioning problem. 
We aim to track the relative position of a group of mobile users by integrating the inertial measurement provided by the IMU and the peer to peer ranging measurement from the UWB. 
As our approach does not rely on any given infrastructure, 
the environment is only used to validate our approach and provide the ground truth.
User1, User2, and User4 can range each other using UWB, since they are close to each other.
Due to the occlusion of the concrete walls, User3 can not hear User1 through UWB.
With the help of IMU, we can cooperatively determine the relative positions of all users.
}
\label{fig:example}
\vspace{-0.2in}
\end{figure}

In some scenarios, for example fire rescue within a building, the global positioning is not possible, since anchors may not be deployed or not functional due to the accident.
Therefore, relative positioning of users without any external infrastructure is appealing, which is the focus of this paper. 
In the context of relative positioning as shown in Fig. \ref{fig:example}, 
all users are considered as equal peers and are able to obtain the range information of its neighbors if they are in communication range. 
Additionally, the users carry inertial sensors, which can measure their own movements.
The goal is to determine the relative position of all users in the network. 

Dead reckoning (DR)\,\cite{IMU_Richard}\,\cite{richard_ieeesensor_2016} determines one's location based on its previous position and speed, 
which is measured by an IMU sensor or wheel encoder in the case of a mobile robot. 
If the initial locations of the users are known, one can use DR to determine the relative position of a group of users.
But the DR may not be accurate due to accumulative error, 
which must be corrected or eliminated by other sources of information.
Anchors with known positions, for example, can provide 
a measure to correct the positioning error, 
but infrastructure-based anchors are not applicable in a number of situations as mentioned previously. 
In this paper, we propose using the peer to peer measurement to remove the accumulative error for relative positioning estimation.

Many devices are able to provide peer to peer information, for example camera, laser range finder, and wireless sensors.
Extensive research concerning relative positioning in a swam behavior using vision or laser-based sensors have been done in the area of robotics \cite{Huang201552} \cite{Leccese_swarm_icra_2013}.
The application of these approaches are limited in uncontrolled environments, as it is challenging for them to deal with the occlusions.

Due to the wide availability of RSS in many wireless devices, 
a number of model-based or fingerprinting-based techniques\,\cite{Kafrawy_propagation_modeling}\,\cite{RanArtur_IROS_2012} have been proposed to locate a device. 
As compared to the visual-based sensors mentioned above, the RSS is available even without line of sight. 
But the accuracy of RSS-based approach is limited 
since characterizing radio propagation in an environment is challenging
due to severe multipath and numerous site-specific parameters.
Recently, a novel wireless radio technology called ultra-wideband (UWB)\,\cite{olsson2014cooperative}\,\cite{cooperative_localization_UWB_IMU_soilder} has been widely used to provide ranging information. 
This kind of sensor uses the ToA-based technique to measure the distance traveled and is able to provide a positioning accuracy within a few centimeters, 
which is several times better than RSS-based positioning systems. 

In this paper, we propose an approach to combine the IMU inertial and UWB ranging measurement for relative positioning without any given infrastructure in a probabilistic way.
A dual particle filter is additionally used to incorporate the UWB ranging measure and recover from positioning failures.
On the one hand, the UWB is great at providing the distance information, 
but the communication range is limited and it does not provide any bearing information and may face location ambiguity while using UWB alone for positioning 
as pointed out in our previous work \cite{Liu_Relative_Globecom}. 
In this paper, we show how the motion measurement from IMU can be used to resolve this ambiguity.
On the other hand, the IMU is notorious for the accumulative errors, and we demonstrate how the UWB can be used to remove this kind of error.
As a result, by fusing the measurements from IMU and UWB during a period of time, 
we can take advantages of both sensors and cooperatively estimate the relative positions of mobile users.
In particular, a central server is running in the back-end to fuse all measurements which does not require any computation at the sensor unit. 
Fig.\,\ref{fig:example} illustrates the concept of our relative positioning system. 
Considering the simulation and the real experiments conducted in this paper, 
we believe that the IMU inertial and UWB ranging can be used for cooperative localization in many scenarios, like firefighter operations and searching in disaster areas.

The rest of this paper is structured as follows. 
We review the related literature in Sect.\,\ref{related_work}. 
Sect.\,\ref{system_overview} formulates the problem to be solved, which is followed by the implementation using a particle filter in Sect.\,\ref{particle_filtering_fusion}.
We experimentally validate the above mentioned in Sect.\,\ref{Simulations} and Sect.\,\ref{experimental_evaluations}.
Finally, we draw conclusions in Sect.\,\ref{conclusions}.

\section{Related Work}
\label{related_work}
Over the last decade,
there is a growing interest in indoor positioning due to the rapid demand 
of many location-aware services.
Many researchers focus on the global positioning in given infrastructures. 
For example, the existing WLAN-based infrastructures are usually covered by a number of wireless access points (APs). 
Many off-the-shelf devices (i.e. smart phones) are able to provide the RSS, 
which can be used to infer mobile's locations\,\cite{Yang_using_human_motions}\,\cite{Zero_calibration}. 
In many scenarios, 
the priori knowledge of an infrastructure is not feasible, such as personnel searching in disaster areas. 
Therefore, many researchers focus on relative positioning rather than global positioning. 
Authors in \cite{relative_location_without_gps} presented an algorithm to achieve relative positioning for static sensor nodes in a sensor network.
The location ambiguity exists in some nodes during positioning \cite{Analysis_Flip_Ambiguities}\,\cite{OFA_optimistic_flip_handle} 
and will propagate to other nodes which results in a poor positioning accuracy. 

Due to the mobility of the sensor nodes,
cooperative relative positioning by fusing IMU and ranging information attracts more and more attentions
\cite{olsson2014cooperative}\,\cite{cooperative_localization_UWB_IMU_soilder}\,\cite{Nilsson_2013_iros_cooperative_Initialization}\,\cite{strader2016cooperative}\,\cite{zhou_robot_robot_range_motion_2008}.
Authors in\,\cite{zhou_robot_robot_range_motion_2008} consider the relative positioning of two mobile robots under ideal ranging and motion measurement.
This is reasonable since the motion of the robot can be precisely measured by wheel encoders.
However, this does not apply to mobile user positioning due to irregular movements of the users, e.g. walking sideways and crawling.
Authors in\,\cite{Nilsson_2013_iros_cooperative_Initialization} 
propose an approach to initialize the orientation of an agent based on ranging and dead reckoning.
The initial state of an agent is recursively estimated by considering dead reckoning of the nearby agents. 
However, their approach can not correct the positioning error after the initialization stage. 
Authors in\,\cite{olsson2014cooperative} used an extended Kalman filter to implement a cooperative localization system for firefighters
by fusing UWB and IMU information in a decentralized way. 
The fusion is done in each sensor unit, which additionally requires users to communicate with each other to share its state in order to incorporate the ranging measurements.

\begin{figure}
\centering
\includegraphics[width=0.47\textwidth]{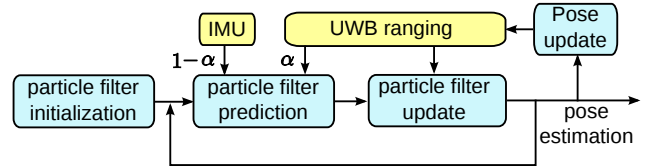}
\caption{Particle filtering for sensor fusion.}
\label{fig:fusion_overview}
\vspace{-0.2in}
\end{figure}

\section{Problem Formulation}
\label{system_overview}
Various wireless devices, i.e. Wifi, UWB, and RFID, provide the raging information, for example RSS and ToF (time of flight), 
which can infer the potential location of a target.
Wireless signals can easily go through obstacles and show a big advantage over the visual-based sensors. 
Recently, a novel wireless technology called UWB is able to provide accurate range information through time of flight.

Assuming an indoor scenario consisting of $N$ mobile users $\x_{1:N}^{(t)}$ with unknown positions at time $t$, 
each user $i$ is capable of measuring the distance of the neighbors $\z_i^{(t)}=\{z_{ij}^{(t)}\}_{j \in \N_i}$, 
where $\N_i$ denotes the set of neighboring users sensed by the $i$th user.
It is important to note that in the case that two users are out of range or being blocked, 
we may not receive any ranging value. 
However, we assume whenever there is a ranging value (i.e. $z_{ij}^{(t)}$), it has high accuracy, and we refer the readers to later section (Sect.\,\ref{similarity_UWB}) for more details on its modeling. 
Moreover, each user $i$ carries an IMU sensor which is able to measure its own movement $\m_i^{(t)}=\Delta \x_i^{(t)}$. 
We aim to determine the two-dimensional positions and orientations of all users in a local coordinate frame without any given reference infrastructure. 

Formally, 
to estimate the unknown positions of users $\x_{1:N}^{(t)}$ at time $t$ given the sequence of ranging and motion measurements, 
which are denoted as $\Z=\{\z_{i}^{(1)},...,\z_{i}^{(t)}\}$ and $\M=\{\m_{i}^{(1)},...,\m_{i}^{(t)}\}$ respectively, 
we need to construct the joint posterior probability $p(\x_{1:N}^{(t)}|\M,\Z)$.
We assume motion and ranging measurements are independent. 
According to Bayesian theory and the Markov assumption, $p(\x_{1:N}^{(t)}|\M,\Z)$ can be factorized into:
 \begin{equation}
 \begin{split}
  &p(\x_{1:N}^{(t)}|\M,\Z) \propto {p(\M,\Z|\x_{1:N}^{(t-1)}) \cdot p(\x_{1:N}^{(0)})}\\
&=\prod_{i=1}^{N} \underbrace{p(\x_{i}^{(t)}|\x_{i}^{(t-1)},\m_i^{(t)})}_\text{IMU Inertial} \cdot \prod_{i=1}^{N} \underbrace{\prod_{j \in \N_i}{p(z_{ij}^{(t)}|\x_i^{(t)},\x_j^{(t)})}}_\text{UWB Ranging}
\cdot p(\x_{1:N}^{(0)}),
 \end{split}
  \label{eq:bayesian_framework}
 \end{equation} 
here 
$p(\x_{1:N}^{(0)})$ is the prior location information of users at $t=0$. 
$p(\x_{i}^{(t)}|\x_{i}^{(t-1)},\m_i^{(t)})$ is the motion model, which predicts the pose of a user $\x_i^{(t)}$ at time $t$ given the previous pose $\x_i^{(t-1)}$ 
and the displacement information from the IMU $\m_i^{(t)}$. 
$p(z_{ij}^{(t)}|\x_i^{(t)},\x_j^{(t)})$ is the ranging model of the UWB measurement, 
which represents the likelihood of receiving a ranging measurement $z_{ij}^{(t)}$ given the states of two users $\x_i^{(t)}$ and $\x_j^{(t)}$. 
The motion model and the ranging model will be detailed in Sect.\,\ref{IMU} and Sect.\,\ref{similarity_UWB} respectively.
\section{State Estimation with the Particle Filtering}
\label{particle_filtering_fusion}

There are many implementations of the recursive Bayesian framework, e.g. particle filters and Kalman filters. 
As a non-parametric implementation of Bayesian framework, 
particle filters approximate the distribution with a collection of samples and has no assumption about the distribution of the probability density function. 
Therefore, we choose particle filters to fuse the measurements from different sources.
An overview of the sensor fusion algorithm is shown in Fig.\,\ref{fig:fusion_overview}.
\subsection{Particle Filtering}
\label{state_estimation}



In particular, the particle filter is represented by a set of $M$ particles $\x_i^{(t)}=\{\x_i^{(t,k)},w_i^{(t,k)}\}_{k=1}^{M}$, 
where $\x_{i}^{(t,k)}=\{x_i^{(t,k)},y_i^{(t,k)},\theta_i^{(t,k)}\}$ is the 2D pose hypotheses and $w_i^{(t,k)}$ is the associated weight.
The pose of the user is computed by a weighted mean among all particles. 
In general, the particle filter is performed recursively with the following three steps: 
\begin{itemize}
 \item \textbf{Prediction:} We draw a new set of particles according to the motion model $p(\x_{i}^{(t)}|\x_{i}^{(t-1)},\m_i^{(t)})$, 
  which is determined by the input of the IMU carried by a user (see Sect.\,\ref{IMU} for more detail). 
   
 \item \textbf{Correction:} We assign each particle with a new weight according to UWB ranging model (Sect.\,\ref{similarity_UWB}) when a new measurement $\z_i^{(t)}$ arrives. 
 \item \textbf{Resampling:} We generate a set of new particles as a replacement of the set of old particles if the effective sample size falls below a threshold $\frac{M}{2}$. 
 In general, the probability that a particle appears in the new particle set depends on its weight. 
\end{itemize}

\begin{figure}
  \centering
    \subfigure[]{
\label{fig:IMU_single}
    \includegraphics[width=0.26\textwidth]{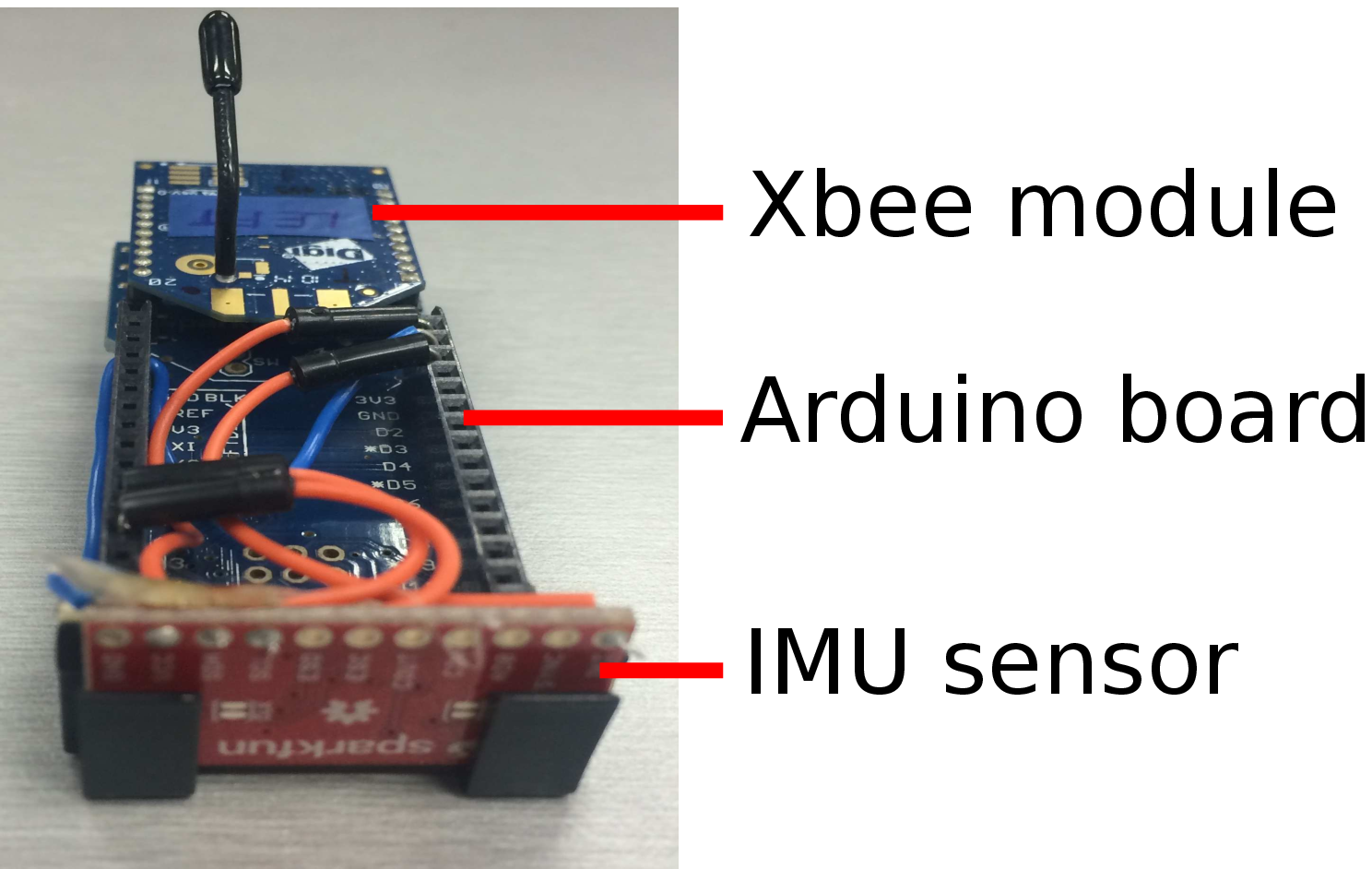}
    }
  \subfigure[]{
\label{fig:IMU_overall}
        \includegraphics[width=0.18\textwidth]{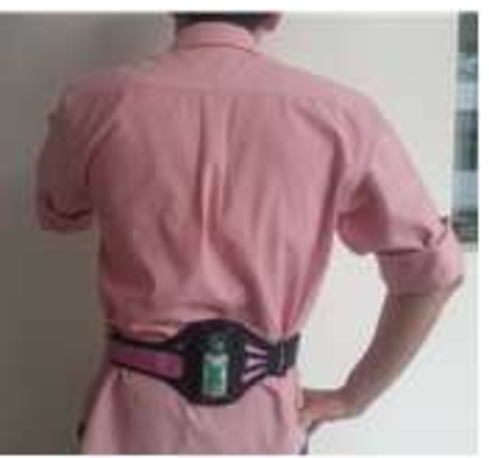}
        }
        \subfigure[]{
\label{fig:UWB_tx}
        \includegraphics[width=0.2\textwidth]{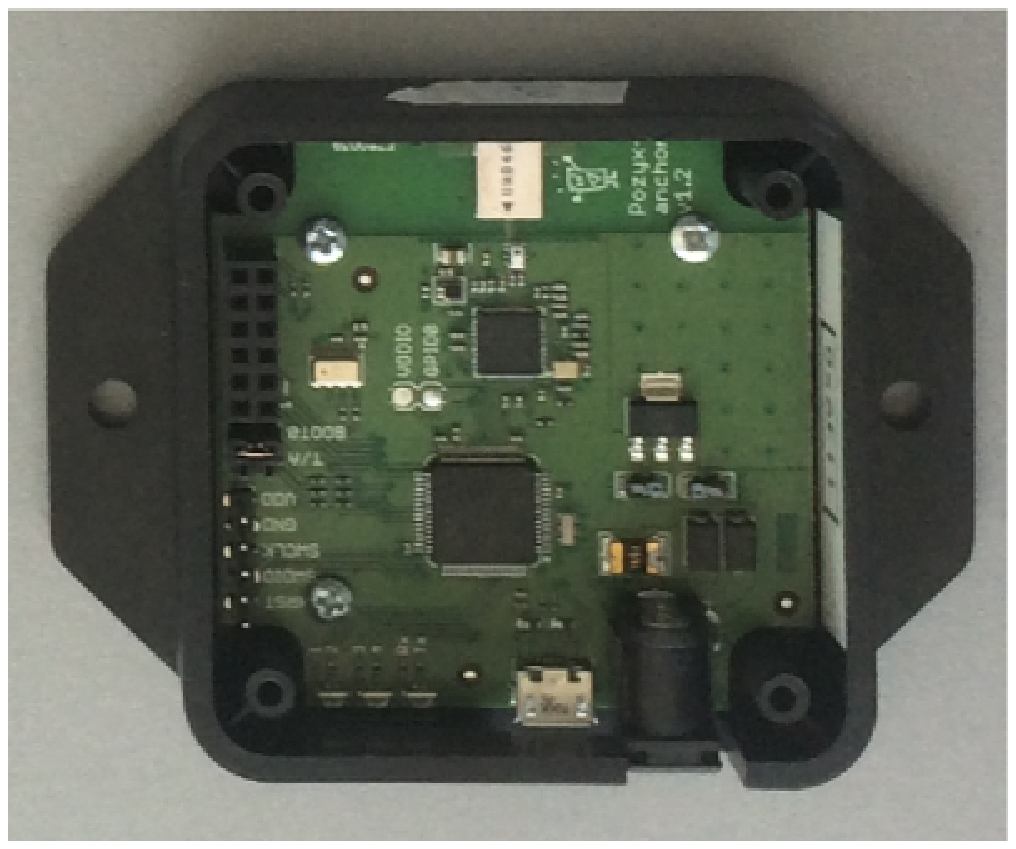}
        }
        \subfigure[]{
\label{fig:UWB_tx}
        \includegraphics[width=0.25\textwidth]{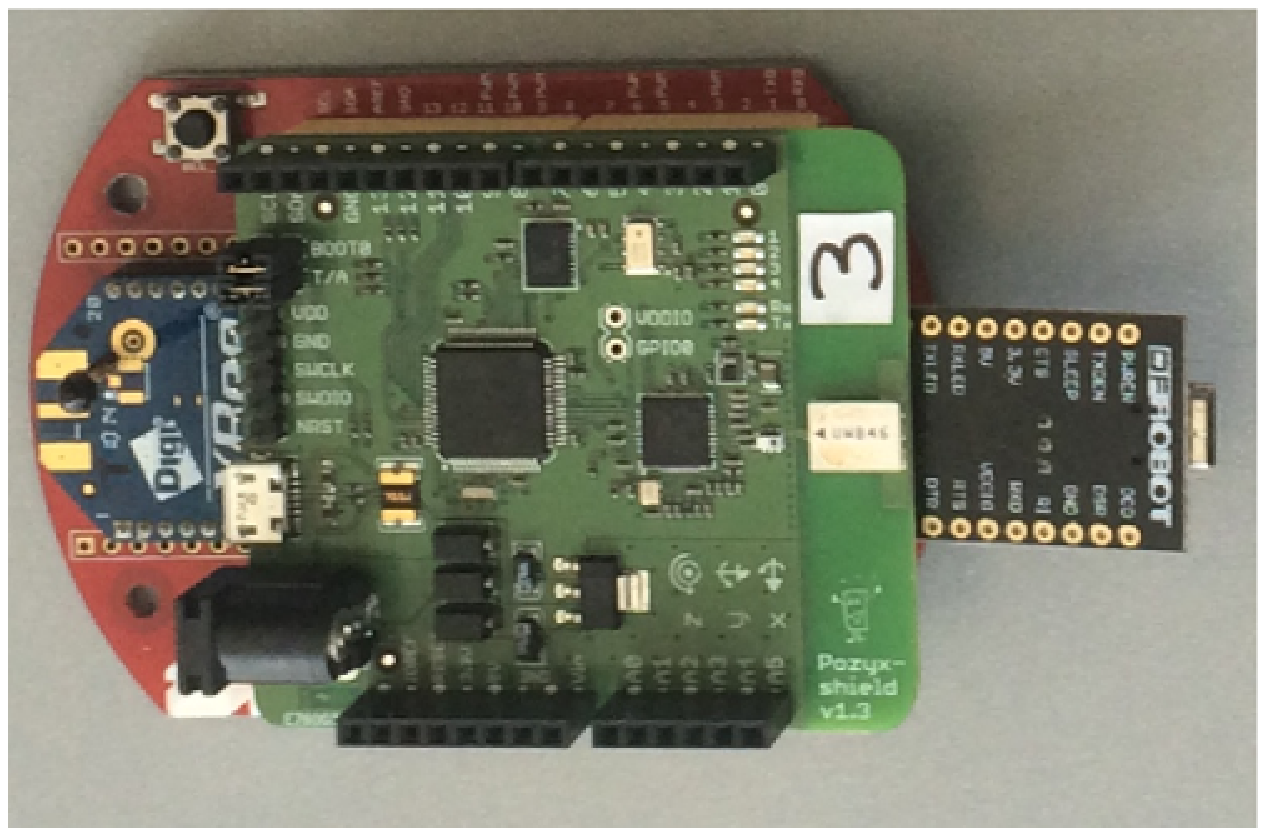}
        }         
   \caption[sensors]
{Sensors carried by a user. (a) IMU sensor; (b) IMU sensor placed on the upper torso of a user; (c) UWB transmitter; (d) UWB receiver.}
\label{fig:sensor}
\vspace{-0.2in}
\end{figure}
  
\subsection{IMU Mounted on Upper Torso for Dead Reckoning}
\label{IMU}
We placed an IMU sensor on the upper torso of a user for dead reckoning as shown in Fig.\,\ref{fig:sensor}. 
The IMU consists of a 3D accelerometer, a 3D gyroscope, and a 3D magnetometer. 
In our previous work \cite{IMU_Richard} \cite{richard_ieeesensor_2016}, 
we placed the IMU sensor at the ankle for dead reckoning,
but the end-users commented that placing the IMU at the ankle affects their walking. 
Thus, we propose to place the IMU sensor on the upper torso (see \cite{richard_ieeesensor_2016} for the details).

In general, we applied the indirect Kalman filter \cite{IMU_Richard} \cite{richard_ieeesensor_2016} to get a smooth estimation of the displacement of a user. 
The proposed method does not require any prior training and the leg length can be estimated using an inverted pendulum model.
As a result, the IMU reports the displacement estimation $\Delta \x_i^{(t)}=(\Delta x_i^{(t)},\Delta y_i^{(t)},\Delta \theta_i^{(t)})$ of user $i$ at time $t$. 
A micro-controller is used to send the displacement estimation (i.e. $\Delta x_i^{(t)}$, $\Delta y_i^{(t)}$, and $\Delta \theta_i^{(t)}$) to the server 
as inputs of the particle filtering (also see Fig.\,\ref{fig:fusion_overview}). 
We predict the state of particles upon the IMU measurement corrupted with a Gaussian noise: 
\begin{align}
x_i^{(t)}=x_i^{(t-1)}+\Delta x_i^{(t)}+\Delta x_i^{(t)} \cdot \mathcal{N}(0, \sigma_d^2)\\
y_i^{(t)}=y_i^{(t-1)}+\Delta y_{i}^{(t)}+\Delta y_i^{(t)} \cdot \mathcal{N}(0, \sigma_d^2)\\
\theta_i^{(t)}=\theta_{i}^{(t-1)}+\Delta \theta_i^{(t)}+\Delta \theta_i^{(t)} \cdot \mathcal{N}(0, \sigma_{\theta}^2)
  \label{eq:motion_model}
 \end{align}
 where $\sigma_d^2$ and $\sigma_\theta^2$ are Gaussian noises added to the distance displacement and orientation respectively.




IMU is quite accurate at estimating the change of position over short periods of time, 
but is very sensitive to error accumulation over long term run. 
In particular, for a mobile user, the heading is highly influenced by the irregular movements and the local magnetic field disturbances in the indoor environment. 
Therefore, we utilize the ranging measurements from UWB to compensate for the errors in IMU to generate a new, more accurate, and reliable relative positioning system.

\subsection{Ranging Model of UWB Measurement}
\label{similarity_UWB}

We assume the noise from the range measurement is Gaussian with a standard deviation $\sigma_r$:
\begin{equation}
\label{eq:similarity_measure}
p(z_{ij}^{(t)}|\x_{i}^{(t)},\x_{j}^{(t)})=\mathcal{N}(\left \| \x_{i}^{(t)}-\x_{j}^{(t)} \right \|, \sigma_r^2)
\end{equation}

Therefore, the likelihood of receiving a ranging measurement $z_{ij}^{(t)}$ given the states of the two nodes $\x_i^{(t)}$ and $\x_j^{(t)}$ is computed as:
\begin{equation}
\label{eq:similarity_measure}
p(z_{ij}^{(t)}|\x_{i}^{(t)},\x_{j}^{(t)})=\frac{1}{\sqrt{2\pi}\sigma_{r}}\exp \left (-\frac{{(z_{ij}^{(t)}-d(\x_i^{(t)},\x_j^{(t)}))^{2}}}{2\sigma_{r}^2} \right)
\end{equation}
where $d(\cdot)$ is the square root distance between two estimations. 
In this paper, we only consider positive ranging measurement. 
In general, negative detection is usually considered to be less useful than positive information (see \cite{Hoffmann06_negative}).
For example, detecting a user provides much more information than not observing a user, since there are many potential positions where one user is not able to detect the other user.
In our experiment, even if two users are very close, it is still possible that one user can not communicate with the other user due to non-line-of-sight effect.



\subsection{Dual Particle Filter to Recover from Positioning Failures}
\label{sect:dual_MCL}

Due to the irregular movements of users, it is very hard to find a universal model to feature the error characteristics of the IMU. 
As a result, the particle filter may place a small number of particles (or no particles at all) around the true pose of the target, which leads to positioning failures. 
To solve this issue, authors in \cite{Lenser00sensorresetting} proposed sensor resetting, which adds new samples according to the current measurement likelihood.
Authors in \cite{Gutmann_compare_MCL_IROS2012} proposed another way to determine the number of particles to be added based on two smoothed estimations of measurement likelihoods.
But the newly added samples may introduce an inconsistency to the current probability density function. 
This paper solves this problem using the dual particle filter~\cite{Thrun00montecarlo}, which adds particles based the current measurement and determines their weights based on the current probability density function. 
For our application, it is straightforward to draw particles based on the current measurement (i.e. ranging $z_{ij}^{(t)}$), 
as we assume the ranging measurement by UWB is precise. 

Particularly, the importance weights of newly added particles are determined by reconstructing the belief using kernel density estimation (KDE) (see~\cite{ram2011density} in detail) based on the current state estimation. 
Therefore, we draw $\alpha$ samples using the dual particle filter based on the current measurement, $1-\alpha$ samples according to the motion model from IMU (see Fig. \ref{fig:fusion_overview}):
\begin{equation}
\begin{split}
\label{eq_rss_likelihhod}
 &\x_{i}^{(t)} \sim \alpha \cdot \frac{p(\z_i^{(t)}|\x_{i}^{(t)})}{\pi(\z_i^{(t)},\x_{i}^{(t)})}+(1-\alpha) \cdot {p(\x_{i}^{(t)}|\x_{i}^{(t-1)},\m_i^t)},
\end{split}
 \end{equation}
where $0\leq \alpha \leq1$ and $\pi(\z_i^{(t)},\x_{i}^{(t)})=\int p(\z_i^{(t)}|\x_{i}^{(t)}) \mathrm{d}\x_{i}^{(t)}$. 
We refer the readers to \cite{Thrun00montecarlo} for details of the implementation.
\begin{figure}
  \centering
    \subfigure[]{
\label{fig:simulation}
    \includegraphics[width=0.45\textwidth]{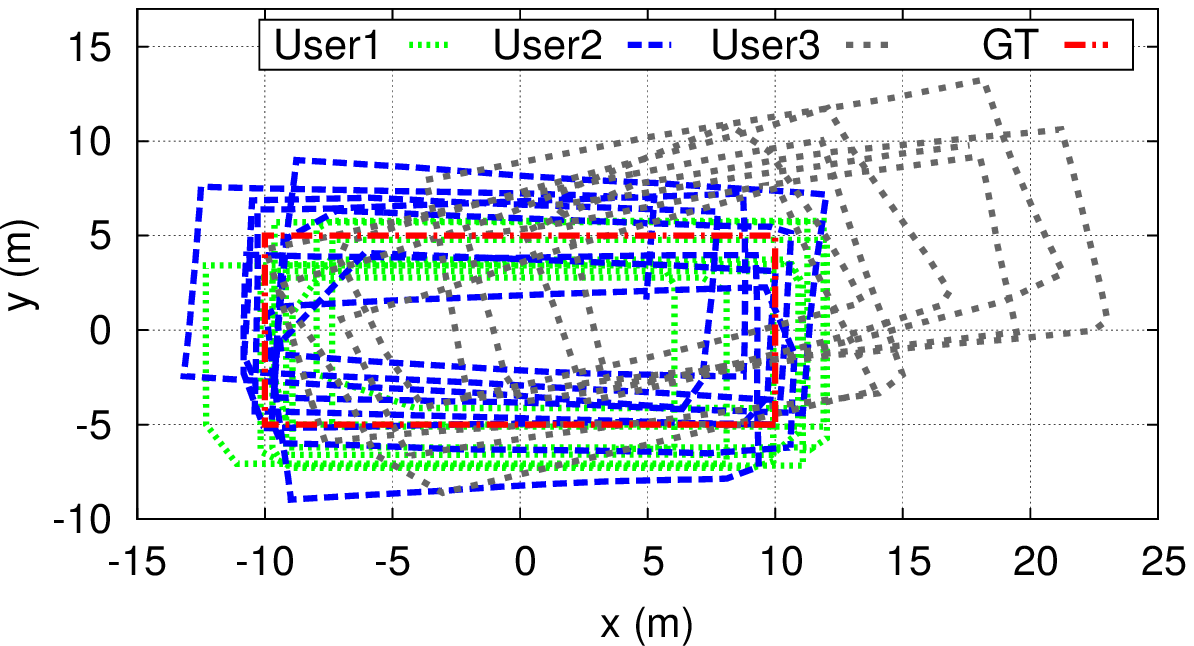}
    }
  \subfigure[]{
\label{fig:simulation_estimation}
        \includegraphics[width=0.45\textwidth]{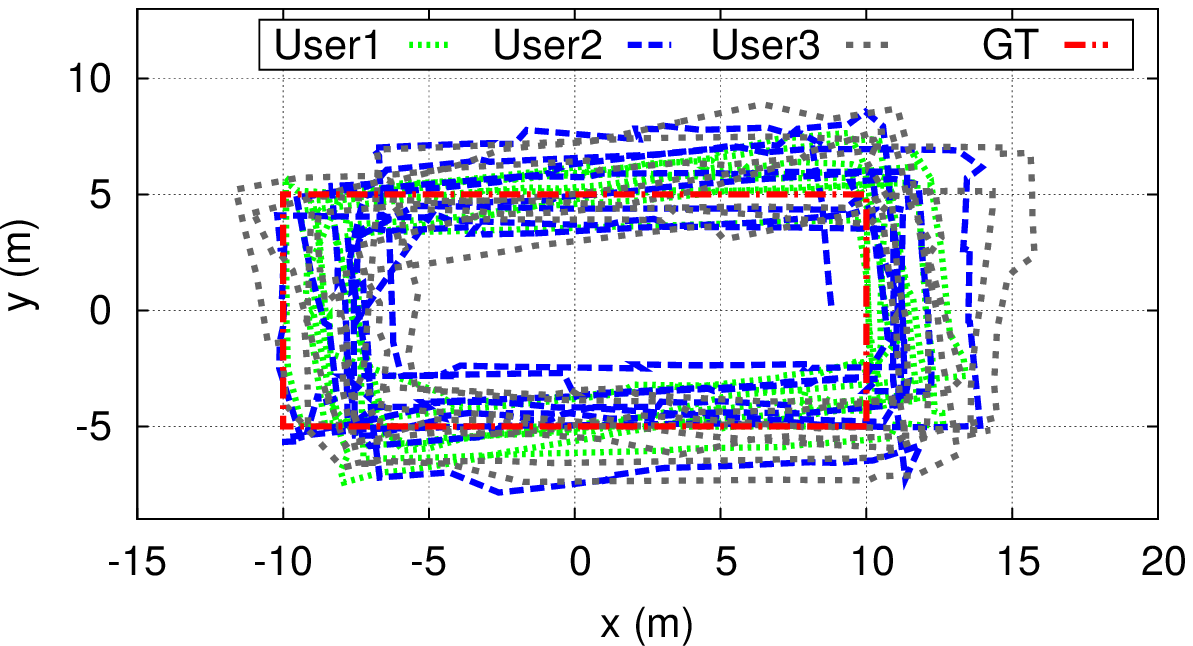}
        }
   \caption[Simulations Experiments]
   {The IMU track (Top) and the track by incorporating UWB (Bottom) ranging measurements in the simulation. 
   User1 is added with a small IMU noise and User3 with a larger IMU noise.}   
\label{fig:simulation_experiments_track}
\vspace{-0.2in}
\end{figure}

\section{Simulations}
\label{Simulations}
We first evaluated our approach in a simulation in this section and then validated the approach in a real world experiment in Section \ref{experimental_evaluations}.
The goal of the simulation is to evaluate the accuracy and robustness of our approach.
Moreover, the simulation gives a thorough investigation of the key parameters and help to choose the best parameters.

\subsection{Simulations Setups}
\label{Simulations setups}
We generated a scenario consisting of six users walking along a rectangle path multiple times in an environment of 20\,m$\times$10\,m, as shown in Fig. \ref{fig:simulation}.
All users started from the same location and kept a distance of approx. 8 meters during the walking.
The speed of the user is about 0.5 m/s.
We produced UWB ranging and IMU inertial measurement with a frequency of 0.5 HZ and 1 HZ respectively. 
For different users, various Gaussian noises were added to the step displacements and the heading changes, 
since in actual scenario the user track can be very different based on the individuals (see \cite{olsson2014cooperative}).
The root mean square error (RMSE) of the relative distance among all pairs of users is used as a measure of the mean positioning accuracy.
The server used for sensor fusion is running on an Intel Core i5-4200M\,@\,2.5GHz CPU, with 4GB RAM. 

\subsection{Impact of the Ranging Noise $\sigma_r$ and IMU Noise Scale}
\label{experimental_noise_pozyx_sensor_and_wifi_scale}

\begin{figure*}
  \centering
    \subfigure[]{
\label{fig:positioning_error_with_UWB}
    \includegraphics[width=0.32\textwidth]{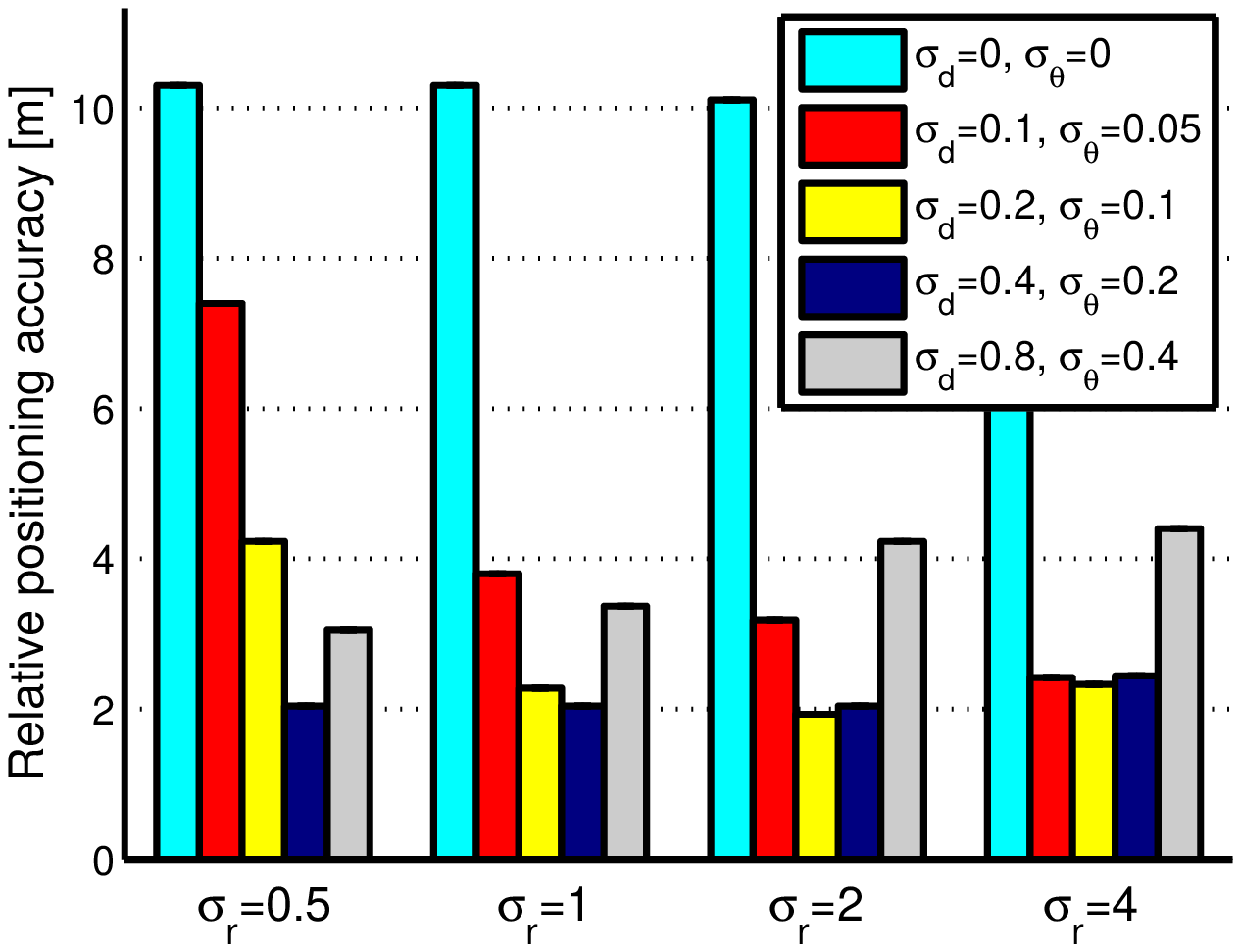}
    }
  \subfigure[]{
\label{fig:particel_size}
        \includegraphics[width=0.30\textwidth]{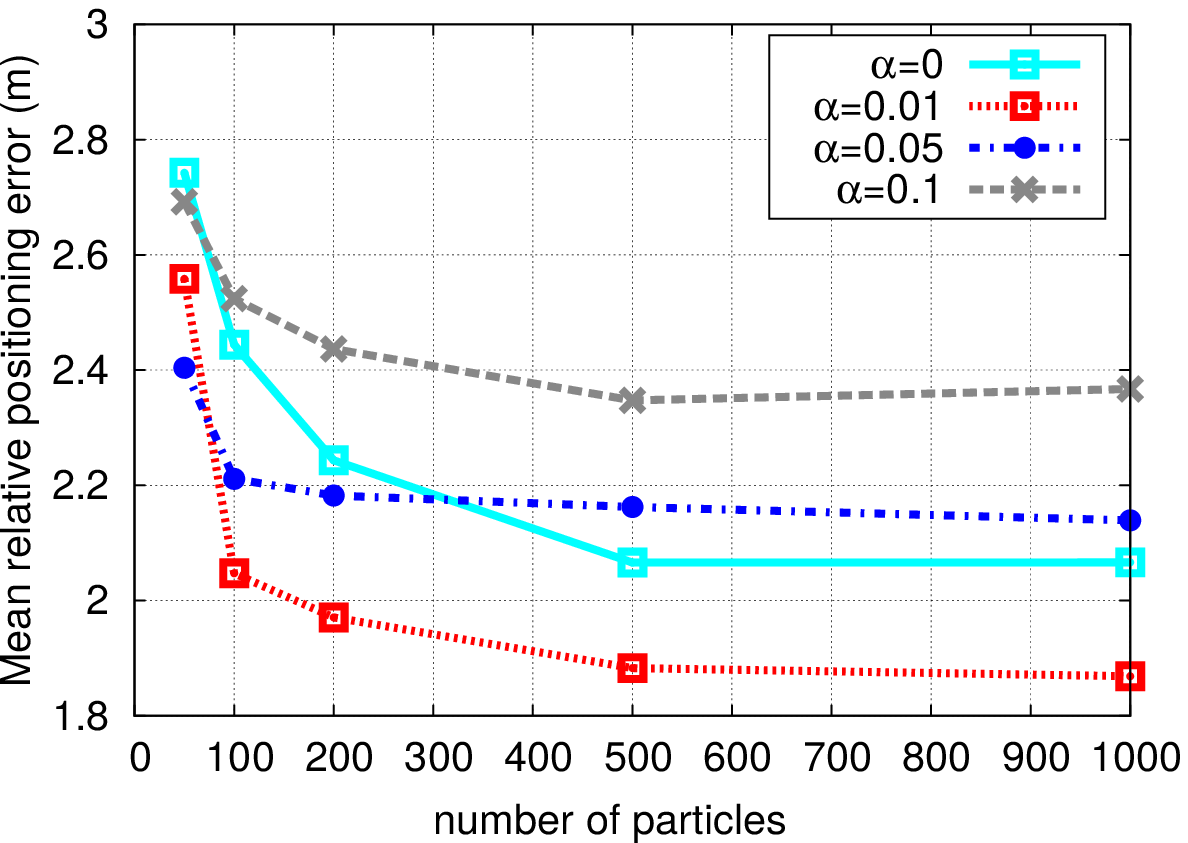}
        }
        \subfigure[]{
\label{fig:positioning_error_with_unknown_init}
        \includegraphics[width=0.32\textwidth]{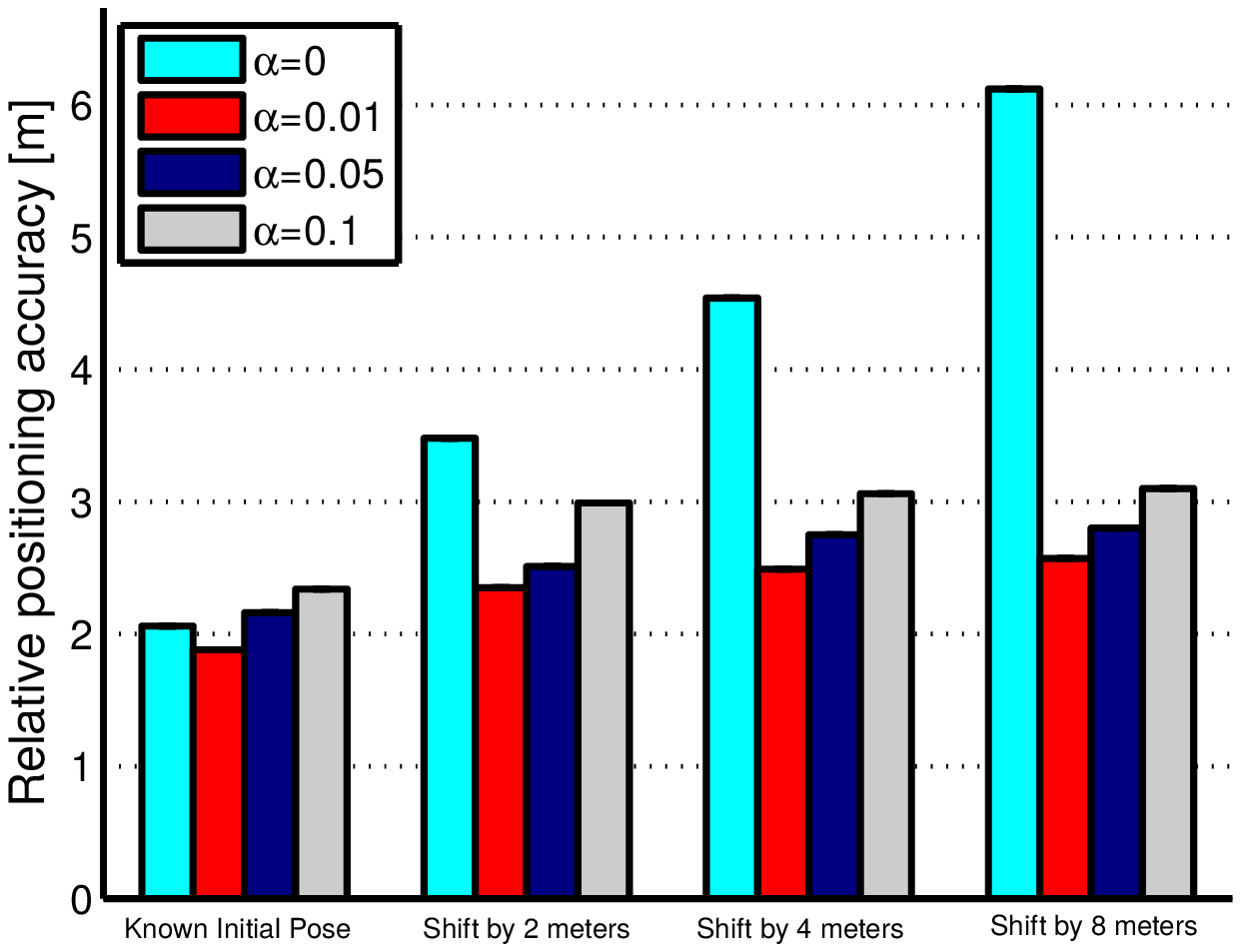}
        }
   \caption[Simulations Experiments]
   {Experiments results in the simulation. (a) Positioning error under the impact of UWB ranging noise $\sigma_r$ and IMU noise (i.e. displacement noise $\sigma_d$ and orientation noise $\sigma_\theta$;
   (b) Impact of the number of particles $M$ and the dual particle filter;
   (c) Positioning error under the impact of wrong initial positions.
   }   
\label{fig:simulation_results}
\vspace{-0.2in}
\end{figure*}

\begin{figure}
\centering
\includegraphics[width=0.45\textwidth]{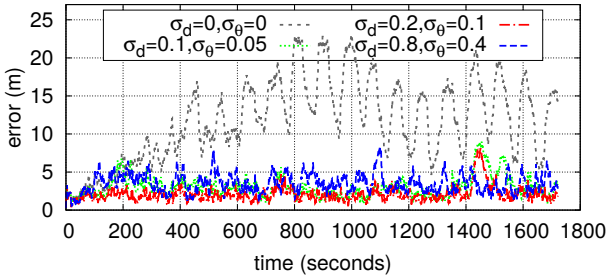}
\caption{
Positioning error over all users in the simulation at different timestamps under the impact of noise added to the IMU.}
\label{fig:simulation_error_different_timestamps}
\end{figure}

We evaluated the positioning accuracy under the impact of different noise scales of UWB sensor $\sigma_r$ 
and various noises added to IMU. 
We assume all users started from the same location, therefore the initial states of particle filters are known. 
We set five different scales of IMU noise, 
i.e. $\{\sigma_d,\sigma_\theta\}$ with the following values:
$\{0,0\}$,
$\{0.1,0.05\}$,
$\{0.2,0.1\}$,
$\{0.4,0.2\}$,
$\{0.8,0.4\}$. 
$\{0,0\}$ can be considered as the case with IMU alone. 
In this series of experiments, we set $\alpha=0.01$ and the number of particles $M=500$. 
Fig.\,\ref{fig:positioning_error_with_UWB} shows the positioning accuracy under different values of $\sigma_r$ and different scales of IMU noise. 
As compared to the accuracy of IMU alone, our approach is more precise. 
For example, we get a mean positioning error of 2.1 m with $\sigma_r=2$ and $\sigma_d=0.2,\sigma_\theta=0.1$, 
which is an improvement by a factor of 5 as compared to the case without UWB (10 m for $\sigma_d=0$ and $\sigma_\theta=0$).
Due to the cumulative characteristic of IMU, 
the relative positioning accuracy will even get far worse for longer tracks. 
Fig. \ref{fig:simulation_error_different_timestamps} plots the positioning error at different timestamps with respect to different IMU noises.  

In general, a larger $\sigma_r$ leads to a worse result. 
This is because a too large noise level will introduce too much noise to the ranging model and results in an unstable estimation, 
thus giving a bad accuracy. 
On the other hand, a too small $\sigma_r$ also leads to a bad positioning result. 
This is because with a too small $\sigma_r$, the particle filter is not able to capture the noise from the UWB sensor, 
and may place a small number of particles or no particles around the true position of the target, which leads to a poor positioning accuracy.

\subsection{Impact of Number of Particles and Dual Particle Filter}
\label{experimental_number_of_particles}

\begin{table}
\centering
\caption{Analysis of the running time (in seconds) under the impact of different number of particles ($M$).}
\label{table:different_particles}
\centering
\begin{tabular}{|c|c|c|c|c|c|}
\hline
 \multirow{1}{*}{\centering  $M$}  &
  \multirow{1}{*}{\centering 50 }&
 \multirow{1}{*}{\centering 100 }&
 \multirow{1}{*}{\centering 200 }&
 \multirow{1}{*}{\centering 500 } &
 \multirow{1}{*}{\centering 1000}\\
\hline
Running time &0.006& 0.012 & 0.026 &0.05& 0.13\\ 
\hline
\end{tabular}
\label{table:tracking_errors_under_different_particles}
\vspace{-0.1in}
\end{table}

Next, we examined the positioning accuracy under the impact of number of particles and different configurations of the dual particle filter as shown in Fig. \ref{fig:particel_size}. 
In this series of experiments, we set $\sigma_r=2.0$ for the UWB ranging model.
For the IMU, we choose $\sigma_d=0.2$ and $\sigma_\theta=0.1$.
We also showed the running time under different number of particles $M$ in Table \ref{table:tracking_errors_under_different_particles}. 
As can be seen from Fig. \ref{fig:particel_size}, the positioning accuracy gets worse with smaller $M$ (e.g. $M\leq200$). 
With $M\geq500$, we achieved nearly the same accuracy. Obviously, the mean computational time required for larger $M$ increases due to the increasing number of particles. 
Integrating one measurement with a particle filter ($M=500$ for example) only requires 0.05 seconds, 
which satisfies the requirement of real-time processing. 
However, the running time will increase if the number of mobile users is increasing. 
Fig. \ref{fig:simulation_estimation} is the estimated track using our cooperative positioning approach with a particle size $M=500$ and $\alpha=0.01$.
As can be seen form this figure, the tracks of all users are aligned with the correction of UWB.

In addition, $\alpha=0.01$ gives the best positioning result, as can be seen from Fig. \ref{fig:particel_size}.
A too large or too small $\alpha$ obviously leads to bad results. 
With $M=500$, 
we achieve a positioning accuracy of 1.8 m,
which is an improvement of 10\% as compared to the case (i.e. $\alpha=0$) without using dual particle filter (2.0 m).
The improvement with the dual particle filter is not significant, as the initial positions of all users are assumed to be known.
In order to show the benefits of the dual particle filter, 
we initialize the particle filter based on a position which is randomly shifted by a certain distance from the true position. 
The mean positioning accuracy is shown in Fig. \ref{fig:positioning_error_with_unknown_init}. 
As can be seen from this figure, 
the accuracy is decreasing due to the wrong initial locations of the particle filter. 
For a shift of 4 meters, we obtain a positioning accuracy of 2.6 m, 
which gives an improvement of 44\% as compared to the case without dual particle filter (4.5 m).
This is because the dual particle is able to place the particle to the true position and deal with the positioning failures, which results in an improved accuracy. 

\section{Real World Experiments}
\label{experimental_evaluations}
\subsection{Implementation Detail}
\label{Implementation Detail}
We used the Pozyx sensor\footnote[1]{{https://www.pozyx.io/}} as UWB module to get the range information. 
Each node has a transmitter and a receiver (see Fig. \ref{fig:sensor}) in order to get the peer-to-peer ranging information. 
The transmitter or receiver has a unique ID which can be used to identify a person.
The Pozyx sensor has a reading range up to 30 meters in clear line-of-sight. 
But the reading range is limited in indoor environments due to the occlusions. 
The sensing data is read by an Arduino board and sent to the server through a Xbee wireless module. 

For the IMU, we used a 9 DOF (degree of freedom) MPU-9150 from SparkFun\footnote[2]{{https://www.sparkfun.com/}}. 
The sensing data is read by an Arduino board via I2C protocol. 
The IMU samples the readings from gyroscope, accelerometer, and magnetometer, fuses them using a pendulum model and an extended Kalman filter \cite{IMU_Richard}, and outputs the displacement information. 
All the processing is done in the Arduino board with a frequency of 50 Hz. 
An XBee wireless module is used to send the computed results to a server for further fusion with UWB ranging measurements.
\begin{figure}
\centering
\includegraphics[width=0.35\textwidth]{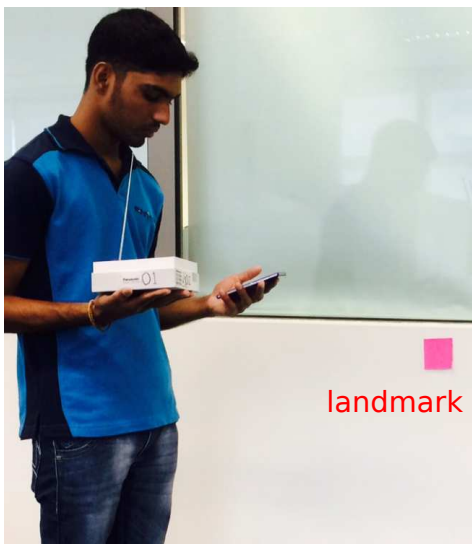}
\caption{A snapshot of the experiment.}
\label{fig:experimental_snapshot}
\vspace{-0.15in}
\end{figure}

\subsection{Evaluation}
\label{evaluation_realworld}
We evaluated the performance of our approach in the laboratory at our campus with a size of 25\,m$\times$15\,m, as shown in Fig.\,\ref{fig:example}. 
During our experiment, three persons carrying UWB sensors (i.e. transmitters and receivers) and IMU sensors (see Fig. \ref{fig:sensor}) walked along a rectangle path multiple times with a normal speed.
The UWB sensor is programmed to send the ranging measurements every 2 seconds.
Although the IMU works at a high frequency (50 Hz), IMU sends the displacement information every one second due to the limitation of Wifi network capacity. 
In total, each person traveled approx. 310 m in 380 s with an average velocity of around 0.8 m/s. 
The resulted track consists of approx. 380 IMU inertial and 190 UWB ranging measurements. 
To record the ground truth, we placed 112 visual landmarks uniformly on the walls. 
When users passed by the landmarks, they are asked to press a button on the mobile phone, 
which will send the ID of the landmark and the timestamp to the server.
The positions of these landmarks are measured before using a Fluke 411D distance meter. 
A snapshot of the experiment is shown in Fig. \ref{fig:experimental_snapshot}. 

\begin{figure}
  \centering
    \subfigure[]{
\label{fig:real_imu_track}
    \includegraphics[width=0.45\textwidth]{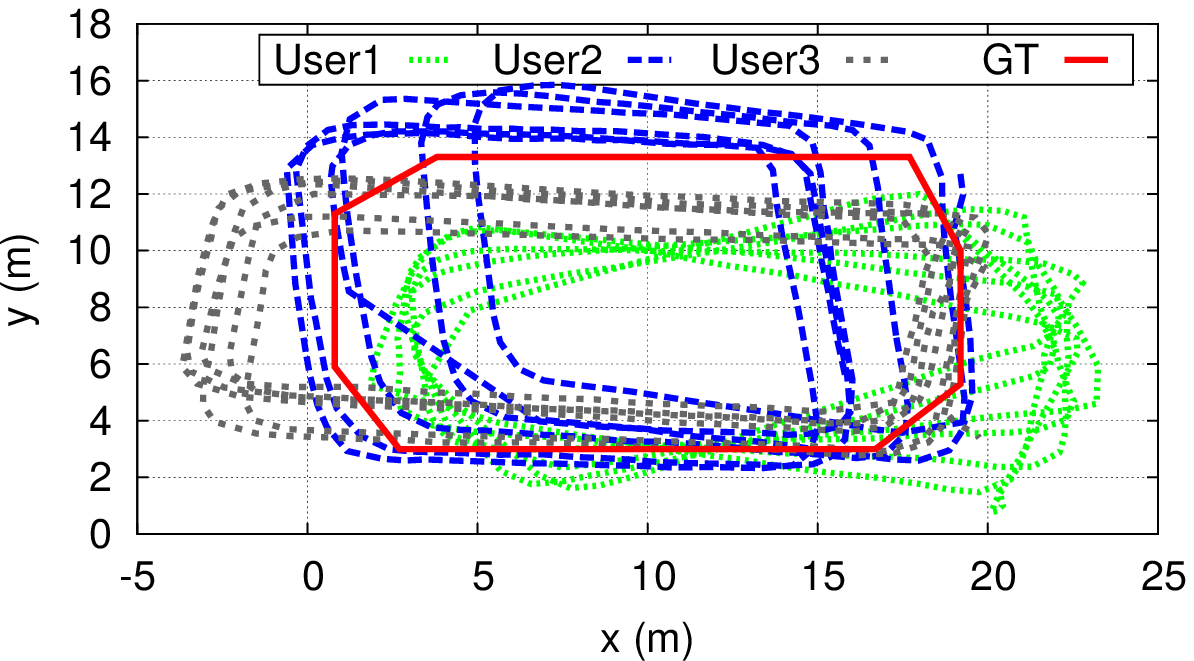}
    }
  \subfigure[]{
\label{fig:real_imu_uwb_track}
        \includegraphics[width=0.45\textwidth]{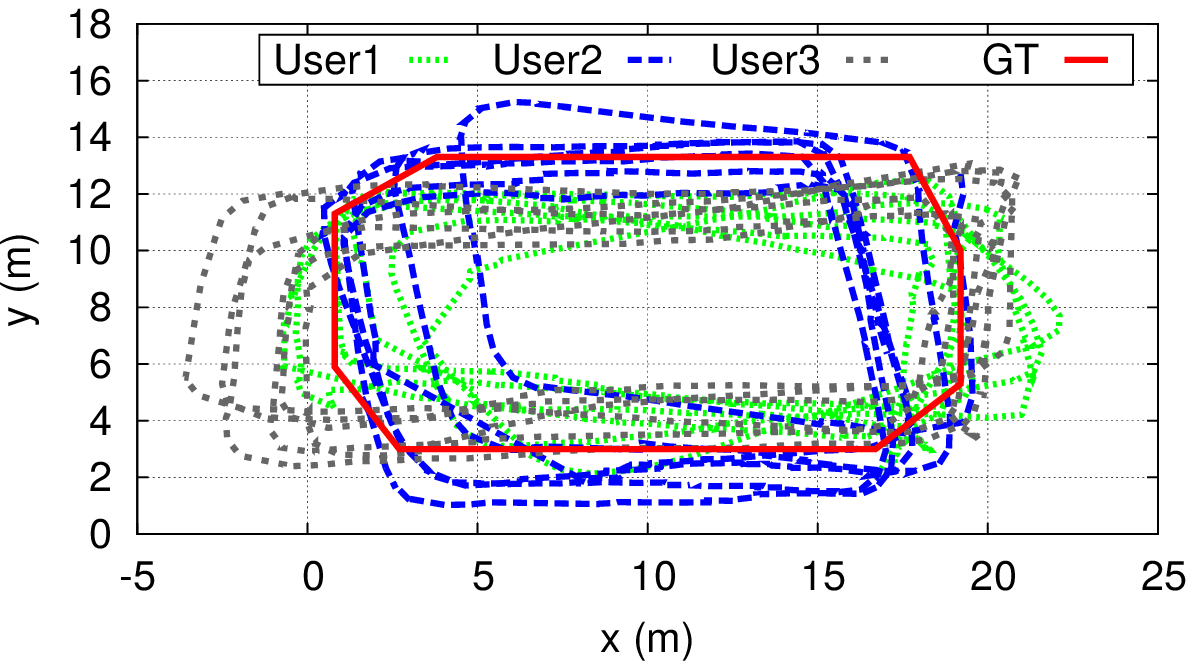}
        }
   \caption[Real Experiments]
   {The IMU track (Top) and the track by incorporating UWB ranging measurements (Bottom) in a real environment. 
   As can be seen from this figure, the IMU tracks of different individuals can be very different.}
\label{fig:real_experiments_track}
\vspace{-0.15in}
\end{figure}

\begin{figure}
\centering
\includegraphics[width=0.49\textwidth]{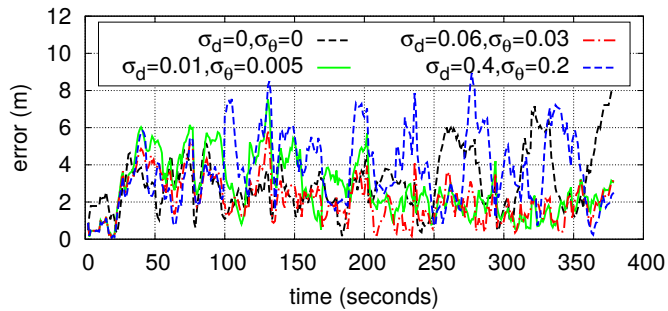}
\caption{Positioning error over all users at different timestamps under the impact of different IMU noises in the real world experiment.}
\label{fig:real_experiments_error_time}
\vspace{-0.15in}
\end{figure}

\begin{figure}
\centering
\includegraphics[width=0.49\textwidth]{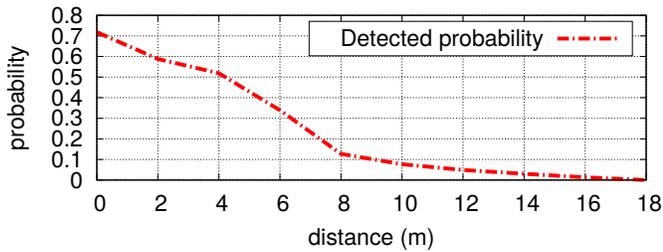}
\caption{Detected probability at different distances during the real world experiments.}
\label{fig:real_detection_probability}
\vspace{-0.15in}
\end{figure}

The original IMU track and track estimated by integrating UWB measurements are shown in Fig. \ref{fig:real_experiments_track}. 
We choose the number of particles $M=500$ and set $\alpha=0.01$. 
We also fixed $\sigma_r=2.0$ for all experiments. 
The relative positioning error over all pairs of users under the impact of different IMU noises are shown in Fig. \ref{fig:real_experiments_error_time}. 
As compared to the raw IMU error (3.0 m), 
we achieve a relative positioning accuracy of 2.2 meters with 3 users, which is much worse than the simulation,
as here for our actual scenario the detection probability of the UWB is limited due to many occlusions in the environment.
We show the detection statistics (i.e. detection probability at different distances) of the UWB sensor in Fig. \ref{fig:real_detection_probability} during the experiment. 
As can be seen from this figure, even if two users are close to each other, there is still some probability that one can not hear another.

\section{Conclusions and Future Work}
\label{conclusions}
We propose an approach to combine the IMU inertial and the UWB ranging information for relative positioning in a probabilistic way without any given infrastructure. 
UWB has very good positioning accuracy, but the communication range is limited due to the occlusion of the environment. 
In contrast, IMU can give a measure of the relative movement of a user, but suffers from the accumulative errors.
Therefore we fuse the measurements from both sensors to compensate the error of an individual sensor and achieve a better positioning accuracy.
A simulation is setup to show the effective of our approach and the parameters are validated through the simulation.
We implemented our approach in a real scenario for multiple users relative positioning and evaluated the performance of our system through experiments.

Our solution is based on the commercially available products and can be further integrated into a single device suitable for many applications, 
such as autonomous mobile robots as well as sensor networks.
We believe the two sensor can be further integrated into a single sensor to position a group of mobile users or agents for the robotics community.
In the future, we would like to extend our approach into 3D and integrate the yaw information from smart phones. 
Another direction is to improve the accuracy of the IMU itself in order to improve the overall relative positioning accuracy.

\bibliographystyle{IEEEtran}
\bibliography{literatur}

\begin{thebibliography}{10}
\providecommand{\url}[1]{#1}
\csname url@rmstyle\endcsname
\providecommand{\newblock}{\relax}
\providecommand{\bibinfo}[2]{#2}
\providecommand\BIBentrySTDinterwordspacing{\spaceskip=0pt\relax}
\providecommand\BIBentryALTinterwordstretchfactor{4}
\providecommand\BIBentryALTinterwordspacing{\spaceskip=\fontdimen2\font plus
\BIBentryALTinterwordstretchfactor\fontdimen3\font minus
  \fontdimen4\font\relax}
\providecommand\BIBforeignlanguage[2]{{%
\expandafter\ifx\csname l@#1\endcsname\relax
\typeout{** WARNING: IEEEtran.bst: No hyphenation pattern has been}%
\typeout{** loaded for the language `#1'. Using the pattern for}%
\typeout{** the default language instead.}%
\else
\language=\csname l@#1\endcsname
\fi
#2}}

\bibitem{Yassin_ieee_tutorials_2016}
A.~Yassin, Y.~Nasser, M.~Awad, A.~Al-Dubai, R.~Liu, C.~Yuen, and R.~Raulefs,
  ``Recent advances in indoor localization: A survey on theoretical approaches
  and applications,'' \emph{IEEE Communications Surveys Tutorials}, vol.~PP,
  no.~99, pp. 1--21, November 2016.

\bibitem{time_of_arrival_2011}
C.~R. Comsa, A.~Haimovich, S.~Schwartz, Y.~Dobyns, and J.~A. Dabin, ``Source
  localization using time difference of arrival within a sparse representation
  framework,'' in \emph{International Conference on Acoustics, Speech and
  Signal Processing (ICASSP 2011)}, May 2011.

\bibitem{angle_of_arrival_2006}
R.~Peng and M.~Sichitiu, ``Angle of arrival localization for wireless sensor
  networks,'' in \emph{3rd Annual IEEE Communications Society on Sensor and Ad
  Hoc Communications and Networks (SECON 2006)}, September 2006, pp. 374--382.

\bibitem{IMU_Richard}
T.~N. Do, R.~Liu, C.~Yuen, and {U-X. Tan}, ``Design of an infrastructureless
  in-door localization device using an imu sensor,'' in \emph{IEEE
  International Conference on Robotics and Biomimetics (ROBIO)}, Zhuhai, China,
  December 6--9 2015.

\bibitem{richard_ieeesensor_2016}
T.~N. Do, R.~Liu, C.~Yuen, M.~Zhang, and {U-X. Tan}, ``Personal dead reckoning
  using imu mounted on upper torso and inverted pendulum model,'' \emph{IEEE
  Sensors Journal}, vol.~16, no.~21, pp. 7600--7608, November 1 2016.

\bibitem{Huang201552}
G.~Huang, M.~Kaess, and J.~J. Leonard, ``Consistent unscented incremental
  smoothing for multi-robot cooperative target tracking,'' \emph{Robotics and
  Autonomous Systems}, vol.~69, pp. 52 -- 67, 2015.

\bibitem{Leccese_swarm_icra_2013}
A.~Leccese, A.~Gasparri, A.~Priolo, G.~Oriolo, and G.~Ulivi, ``A swarm
  aggregation algorithm based on local interaction with actuator saturations
  and integrated obstacle avoidance,'' in \emph{2013 IEEE International
  Conference on Robotics and Automation (ICRA 2013)}, May 2013, pp. 1865--1870.

\bibitem{Kafrawy_propagation_modeling}
K.~El-Kafrawy, M.~Youssef, A.~El-Keyi, and A.~Naguib, ``Propagation modeling
  for accurate indoor {WLAN RSS}-based localization,'' in \emph{the IEEE 72nd
  Vehicular Technology Conference Fall (VTC 2010-Fall)}, Ottawa, Canada, 6--9,
  September 2010, pp. 1--5.

\bibitem{RanArtur_IROS_2012}
R.~Liu, A.~Koch, and A.~Zell, ``Path following with passive {UHF RFID} received
  signal strength in unknown environments,'' in \emph{Porc. of the 2012
  IEEE/RSJ Int. Conf. on Intelligent Robots and Systems (IROS 2012)},
  Vilamoura, Algarve, Portugal, October 2012, pp. 2250--2255.

\bibitem{olsson2014cooperative}
F.~Olsson, J.~Rantakokko, and J.~Nyg{\aa}rds, ``Cooperative localization using
  a foot-mounted inertial navigation system and ultrawideband ranging,'' in
  \emph{2014 International Conference on Indoor Positioning and Indoor
  Navigation (IPIN 2014)}.\hskip 1em plus 0.5em minus 0.4em\relax IEEE, 2014,
  pp. 122--131.

\bibitem{cooperative_localization_UWB_IMU_soilder}
J.~Rantakokko, J.~Rydell, P.~Strömbäck, P.~Händel, J.~Callmer,
  D.~Törnqvist, F.~Gustafsson, M.~Jobs, and M.~Grudén, ``Accurate and
  reliable soldier and first responder indoor positioning: multisensor systems
  and cooperative localization.'' \emph{IEEE Wireless Commun.}, vol.~18, no.~2,
  pp. 10--18, 2011.

\bibitem{Liu_Relative_Globecom}
R.~Liu, C.~Yuen, T.~N. Do, W.~Guo, X.~Liu, and {U-X. Tan}, ``Relative
  positioning by fusing signal strength and range information in a
  probabilistic framework,'' in \emph{the Second IEEE International Workshop on
  Localization and Tracking: Indoors, Outdoors, and Emerging Networks (IEEE
  LION 2016)}, Washington, USA, December 2016.

\bibitem{Yang_using_human_motions}
Z.~Yang, C.~Wu, and Y.~Liu, ``Locating in fingerprint space: Wireless indoor
  localization with little human intervention,'' in \emph{Proceedings of the
  18th Annual International Conference on Mobile Computing and Networking},
  ser. Mobicom '12, 2012, pp. 269--280.

\bibitem{Zero_calibration}
A.~Rai, K.~K. Chintalapudi, V.~N. Padmanabhan, and R.~Sen, ``Zee: Zero-effort
  crowdsourcing for indoor localization,'' in \emph{Proceedings of the 18th
  Annual International Conference on Mobile Computing and Networking}, ser.
  Mobicom '12, 2012, pp. 293--304.

\bibitem{relative_location_without_gps}
S.~Capkun, M.~Hamdi, and J.~Hubaux, ``Gps-free positioning in mobile ad-hoc
  networks,'' in \emph{the 34th Annual Hawaii International Conference on
  System Sciences ( HICSS-34), January, 2001, Hawaii, {USA}}, 2001.

\bibitem{Analysis_Flip_Ambiguities}
A.~A. Kannan, B.~Fidan, and G.~Mao, ``Analysis of flip ambiguities for robust
  sensor network localization,'' \emph{IEEE Transactions on Vehicular
  Technology}, vol.~59, no.~4, pp. 2057--2070, May 2010.

\bibitem{OFA_optimistic_flip_handle}
X.~Wang, Y.~Liu, Z.~Yang, K.~Lu, and J.~Luo, ``Ofa: An optimistic approach to
  conquer flip ambiguity in network localization,'' \emph{Computer Networks},
  vol.~57, no.~6, pp. 1529--1544, 2013.

\bibitem{Nilsson_2013_iros_cooperative_Initialization}
J.-O. Nilsson and P.~Händel, ``Recursive bayesian initialization of
  localization based on ranging and dead reckoning,'' in \emph{In: , 2013
  IEEE/RSJ International Conference on Intelligent Robots and Systems (IROS
  2013)}, 2013, pp. 1399--1404.

\bibitem{strader2016cooperative}
J.~Strader, Y.~Gu, J.~N. Gross, M.~De~Petrillo, and J.~Hardy, ``Cooperative
  relative localization for moving uavs with single link range measurements,''
  in \emph{2016 IEEE/ION Position, Location and Navigation Symposium (PLANS)},
  2016, pp. 336--343.

\bibitem{zhou_robot_robot_range_motion_2008}
X.~Zhou and S.~Roumeliotis, ``Robot-to-robot relative pose estimation from
  range measurements,'' \emph{IEEE Transactions on Robotics}, vol.~24, no.~6,
  pp. 1379--1393, 2008.

\bibitem{Hoffmann06_negative}
J.~Hoffmann, M.~Spranger, D.~Göhring, and M.~Jüngel, ``Making use of what you
  don’t see: Negative information in markov localization,'' in \emph{the
  IEEE/RSJ Int. Conf. of Intelligent Robots and Systems (IROS 2005)}, August
  2005.

\bibitem{Lenser00sensorresetting}
S.~Lenser and M.~Veloso, ``Sensor resetting localization for poorly modelled
  mobile robots,'' in \emph{Proc. of the 2000 IEEE Int. Conf. on Robotics and
  Automation (ICRA 2000)}, San Francisco, CA, USA, April 24--28 2000, pp.
  1225--1232.

\bibitem{Gutmann_compare_MCL_IROS2012}
J.-S. Gutmann and D.~Fox, ``An experimental comparison of localization methods
  continued,'' in \emph{Proc. of the 2002 IEEE/RSJ Int. Conf. of Intelligent
  Robots and Systems (IROS 2002)}, EPFL, Switzerland, Sept. 30--Oct. 4 2002,
  pp. 454--459.

\bibitem{Thrun00montecarlo}
S.~Thrun, D.~Fox, and W.~Burgard, ``{Monte Carlo} localization with mixture
  proposal distribution,'' in \emph{Proc. of the 17th National Conf. on
  Artificial Intelligence (AAAI-2000)}.\hskip 1em plus 0.5em minus 0.4em\relax
  Austin, Texas: MIT Press, July 30--August 3 2000, pp. 859--865.

\bibitem{ram2011density}
P.~Ram and A.~Gray, ``Density estimation trees,'' in \emph{Proceedings of the
  17th ACM SIGKDD International Conference on Knowledge Discovery and Data
  Mining}.\hskip 1em plus 0.5em minus 0.4em\relax ACM, 2011, pp. 627--635.

\end{thebibliography}

\end{document}